# Characterization of $m$-Sequences of Lengths $2^{2k} - 1$ and $2^k - 1$ with Three-Valued Crosscorrelation

Tor Helleseth and Alexander Kholosha and Geir Jarle Ness
The Selmer Center,
Department of Informatics, University of Bergen
PB 7800
N-5020 Bergen, Norway

July 26, 2018

**Abstract.** Considered is the distribution of the crosscorrelation between $m$-sequences of length $2^m - 1$, where $m = 2k$, and $m$-sequences of shorter length $2^k - 1$. New pairs of $m$-sequences with three-valued crosscorrelation are found and the complete correlation distribution is determined. Finally, we conjecture that there are no more cases with a three-valued crosscorrelation apart from the ones proven here.

# 1  Introduction

Let $\{a_t\}$ and $\{b_t\}$ be two binary sequences of length $n$. The *crosscorrelation function* between these two sequences at shift $\tau$, where $0 \leq \tau < n$, is defined by

$$C(\tau) = \sum_{t=0}^{n-1} (-1)^{a_t + b_{t+\tau}} .$$

If the sequences $\{a_t\}$ and $\{b_t\}$ are the same we call it the autocorrelation.

Sequences with good correlation properties are important for many applications in communication systems. A relevant problem is to find the distribution of the crosscorrelation function (i.e., the set of values obtained for all shifts) between two binary $m$-sequences $\{s_t\}$ and $\{s_{dt}\}$ of the same length $2^m - 1$ that differ by a decimation $d$ such that $\gcd(d, 2^m - 1) = 1$. A survey of some of the basic research on the crosscorrelation between $m$-sequences of the same length can be found in Helleseth [1] and more recent results in Helleseth and Kumar [2] and Dobbertin et. al. [3]. A basis for many applications is the family of Gold sequences with their three-valued crosscorrelation function.

In a recent paper [4], Ness and Helleseth studied the crosscorrelation between an $m$-sequence $\{s_t\}$ of length $n = 2^m - 1$ and an $m$-sequence $\{u_{dt}\}$ of length $2^k - 1$, where $m = 2k$ and $\gcd(d, 2^k - 1) = 1$. Here $\{u_t\}$ denotes the $m$-sequence used in constructing the small family of Kasami sequences [5]. Recall that this family consists of $2^k$ sequences $\{s_t\} + \{u_{t+\tau}\}$ for $\tau = 0, \ldots, 2^k - 2$ plus the sequence $\{s_t\}$, where $s_t$ and $u_t$ are defined in (1) and (2). For the Kasami sequences, the crosscorrelation between $\{s_t\}$ and $\{u_t\}$ takes on only two different values. It is an open problem whether this is possible in other cases. Numerical results show several pairs of $m$-sequences with three-valued crosscorrelation function between $\{s_t\}$ and $\{u_{dt}\}$, where $\gcd(d, 2^k - 1) = 1$ and $k$ is odd. In addition to general results, Ness and Helleseth proved in [4] that the decimation $d = \frac{2^k+1}{3}$ gives a three-valued crosscorrelation distribution and in [6] they proved the same distribution for $d = 2^{(k+1)/2} - 1$ (in both cases $k$ odd is needed). In this paper, we cover all the cases found by computer experiments that lead to a three-valued crosscorrelation distribution and completely determine this distribution. Speaking concretely, the decimation $d$ such that $d(2^l + 1) \equiv 2^i \pmod{2^k - 1}$ for some integer $l$ and $i \geq 0$ with $\gcd(l, k) = 1$ and odd $k$ gives a three-valued crosscorrelation distribution. We conjecture that there are no other three-valued cases but these. This result includes the decimations proved in [4, 6] as a particular case that is obtained assuming $l = 1$ and $l = \frac{k+1}{2}$.

In Section 2, we present preliminaries needed for proving our main result. In Section 3, we analyze zeros of a particular affine polynomial $A_a(v)$. In Section 4, we find the distribution of the number of zeros of a special linearized polynomial $L_a(z)$. These two polynomials play a crucial role in finding the distribution of a new three-valued crosscorrelation function. In Section 5, we determine completely the crosscorrelation distribution of the new three-valued decimation.

## 2 Preliminaries

Let $\mathrm{GF}(q)$ denote a finite field with $q$ elements and let $\mathrm{GF}(q)^* = \mathrm{GF}(q) \setminus \{0\}$. The trace mapping from $\mathrm{GF}(q^m)$ to $\mathrm{GF}(q)$ is defined by

$$\mathrm{Tr}_m(x) = \sum_{i=0}^{m-1} x^{q^i} \ .$$

Let $\mathrm{GF}(2^m)$ be a finite field with $2^m$ elements and $m = 2k$ with $k$ odd. Let $\alpha$ be an element of order $n = 2^m - 1$. Then the $m$-sequence $\{s_t\}$ of length $n$ can be written in terms of the trace mapping as

$$s_t = \mathrm{Tr}_m(\alpha^t) \ . \tag{1}$$

Let $\beta = \alpha^{2^k+1}$, then $\beta$ is an element of order $2^k - 1$. The sequence $\{u_t\}$ of length $2^k - 1$ (which is used in the construction of the well-known Kasami family) is defined by

$$u_t = \mathrm{Tr}_k(\beta^t) \ . \tag{2}$$

In this paper, we consider the crosscorrelation between the $m$-sequences $\{s_t\}$ and $\{v_t\} = \{u_{dt}\}$ at shift $\tau$ defined by

$$C_d(\tau) = \sum_{t=0}^{n-1} (-1)^{s_t + v_{t+\tau}} \ , \tag{3}$$

where $\gcd(d, 2^k - 1) = 1$ and $\tau = 0, \ldots, 2^k - 2$. One should observe that in this setting, by selecting all decimations $d$ with this condition, we cover the crosscorrelation function between all pairs of $m$-sequences having these two different lengths. Using the trace representation, this function can be written as an exponential sum

$$C_d(\tau) = \sum_{t=0}^{n-1}(-1)^{s_t+u_{d(t+\tau)}}$$

$$= \sum_{x\in\mathrm{GF}(2^m)^*}(-1)^{\mathrm{Tr}_m(\alpha^{-\tau}x)+\mathrm{Tr}_k(x^{d(2^k+1)})} .$$

Since the two subgroups of $\mathrm{GF}(2^m)^*$ of order $2^k - 1$ and $2^k + 1$, respectively, only contain the element 1 in common, it is straightforward to see that for any element, say $\alpha^{-\tau} \in \mathrm{GF}(2^m)^*$, there is a unique element $u$, where $u^{2^k+1} = 1$ such that $\alpha^{-\tau}u = a \in \mathrm{GF}(2^k)^*$. Further, distinct values of $\tau = 0, 1, \ldots, 2^k - 2$ lead to distinct values of $a \in \mathrm{GF}(2^k)^*$. Further, note that for any $u$ with $u^{2^k+1} = 1$ we have

$$\sum_{x\in\mathrm{GF}(2^m)^*}(-1)^{\mathrm{Tr}_m(\alpha^{-\tau}ux)+\mathrm{Tr}_k(x^{d(2^k+1)})} = \sum_{x\in\mathrm{GF}(2^m)^*}(-1)^{\mathrm{Tr}_m(\alpha^{-\tau}x)+\mathrm{Tr}_k(x^{d(2^k+1)})} .$$

Therefore, the set of values of $C_d(\tau)+1$ for all $\tau = 0, 1, \ldots, 2^k - 2$ is equal to the set of values of

$$S(a) = \sum_{x\in\mathrm{GF}(2^m)}(-1)^{\mathrm{Tr}_m(ax)+\mathrm{Tr}_k(x^{d(2^k+1)})} \qquad (4)$$

when $a \in \mathrm{GF}(2^k)^*$.

The main result of this paper is formulated in the following corollary that gives a three-valued crosscorrelation function between new pairs of sequences of different lengths. This corollary immediately follows from Theorem 2.

**Corollary 1** *Let $m = 2k$ and $d(2^l + 1) \equiv 2^i \pmod{2^k - 1}$ for some odd $k$ and integer $l$ with $0 < l < k$, $\gcd(l, k) = 1$ and $i \geq 0$. Then the crosscorrelation function $C_d(\tau)$ has the following distribution*

$$\begin{array}{lll} -1 - 2^{k+1} & occurs\ \frac{2^{k-1}-1}{3} & times\ , \\ -1 & occurs\ 2^{k-1} - 1 & times\ , \\ -1 + 2^k & occurs\ \frac{2^k+1}{3} & times\ . \end{array}$$

The result will be proved in a series of lemmas. The outline of the proof is as follows. We have shown that we can write $C_d(\tau) + 1$ for $\tau = 0, 1, \ldots, 2^k - 2$ as an exponential sum $S(a)$ for $a \in \mathrm{GF}(2^k)^*$. In the case when $l$ is even, we can calculate the distribution of this sum directly as an exponential sum $S_0(a)$ and obtain the result. In the case when $l$ is odd, a different approach works. In this case, we need some $r$ being a noncube in $\mathrm{GF}(2^m)$ such that $r^{2^k+1} = 1$ (for instance, we can take $r = \alpha^{2^k-1}$ with $\alpha$ a primitive element of $\mathrm{GF}(2^m)$) and we show that
$$S(a) = (S_0(a) + S_1(a) + S_2(a))/3$$
for three exponential sums $S_0(a)$, $S_1(a)$ and $S_2(a)$ defined by
$$S_i(a) = \sum_{y \in \mathrm{GF}(2^m)} (-1)^{\mathrm{Tr}_m(r^i a y^{2^l+1}) + \mathrm{Tr}_k(y^{2^k+1})} \quad \text{for} \quad i = 0, 1$$
$$S_2(a) = \sum_{y \in \mathrm{GF}(2^m)} (-1)^{\mathrm{Tr}_m(r^{-1} a y^{2^l+1}) + \mathrm{Tr}_k(y^{2^k+1})} \ .$$

We determine $S_0(a)$ exactly in Corollary 2 and find $S_1(a)^2$ (that is equal to $S_2(a)^2$) in Lemma 9. Since $S(a)$ is an integer, we can resolve the sign ambiguity of $S_1(a)$ and $S_2(a)$. In order to determine $S_0(a)$ we need to consider zeros in $\mathrm{GF}(2^k)$ of the affine polynomial
$$A_a(v) = a^{2^l} v^{2^{2l}} + v^{2^l} + av + 1$$
and this is done in Section 3. To determine the square sums $S_1(a)^2$ and $S_2(a)^2$ we need to find the number of zeros in $\mathrm{GF}(2^m)$ of the linearized polynomial
$$L_a(z) = z^{2^{k+l}} + r^{2^l} a^{2^l} z^{2^{2l}} + raz$$
and this task is completed in Section 4.

When finding the complete crosscorrelation distribution we make use of the following result from [4] that gives the sum of the crosscorrelation values as well as the sum of their squares.

**Lemma 1 ([4])** *For any decimation $d$ with $\gcd(d, 2^k - 1) = 1$ the sum (of the squares) of the crosscorrelation values defined in (3) is equal to*
$$\sum_{\tau=0}^{2^k-2} C_d(\tau) = 1 \ ;$$
$$\sum_{\tau=0}^{2^k-2} C_d(\tau)^2 = (2^m - 1)(2^k - 1) - 2 \ .$$

## 3 The Affine Polynomial $A_a(v)$

In this section, we take any $k$ and consider zeros in $\mathrm{GF}(2^k)$ of the affine polynomial
$$A_a(v) = a^{2^l} v^{2^{2l}} + v^{2^l} + av + 1 \ , \tag{5}$$

where $l < k$ is an arbitrary but fixed positive integer with $\gcd(l,k) = 1$ and $a \in \mathrm{GF}(2^k)^*$. Let also $l' = l^{-1} \pmod{k}$. The distribution of the zeros in $\mathrm{GF}(2^k)$ of (5) will determine to a large extent the distribution of our crosscorrelation function.

We need the following sequences of polynomials that were introduced by Dobbertin in [7] (see also [8]):

$$\begin{aligned}
F_1(v) &= v, \\
F_2(v) &= v^{2^l+1}, \\
F_{i+2}(v) &= v^{2^{(i+1)l}} F_{i+1}(v) + v^{2^{(i+1)l} - 2^{il}} F_i(v) \quad \text{for} \quad i \geq 1, \\
G_1(v) &= 0, \\
G_2(v) &= v^{2^l-1}, \\
G_{i+2}(v) &= v^{2^{(i+1)l}} G_{i+1}(v) + v^{2^{(i+1)l} - 2^{il}} G_i(v) \quad \text{for} \quad i \geq 1.
\end{aligned}$$

These are used to define the polynomial

$$R(v) = \sum_{i=1}^{l'} F_i(v) + G_{l'}(v) . \qquad (6)$$

As noted in [7], the exponents occurring in $F_j(v)$ (resp. in $G_j(v)$) are precisely those of the form

$$e = \sum_{i=0}^{j-1} (-1)^{\epsilon_i} 2^{il} ,$$

where $\epsilon_i \in \{0,1\}$ satisfy $\epsilon_{j-1} = 0$, $\epsilon_0 = 0$ (resp. $\epsilon_0 = 1$) and $(\epsilon_i, \epsilon_{i-1}) \neq (1,1)$.

Further, we will essentially need the following result proven in [7, Theorem 5] that the following polynomial

$$D(v) = \frac{\sum_{i=1}^{l'} v^{2^{il}} + l' + 1}{v^{2^l+1}} \qquad (7)$$

is a permutation polynomial on $\mathrm{GF}(2^k)^*$. (To be formally more precise, we get a *polynomial* $D(v)$ if $v^{-(2^l+1)}$ is substituted by $v^{(2^k-1)-(2^l+1)}$.) Moreover, $D(v)$ and $R(v^{-1})$ are inverses of each other [7, Theorem 6], i.e., for any nonzero $x, y \in \mathrm{GF}(2^k)$ with $D(x) = y^{-1}$ it always holds that $R(y) = x$. In (7) and in the rest of the paper, whenever a positive integer $e$ is added to an element of $\mathrm{GF}(2^k)$, it means that added is the identity element of $\mathrm{GF}(2^k)$ times $e \pmod 2$.

Also note the fact that since $l'l \equiv 1 \pmod{k}$ then

$$(2^l - 1)(1 + 2^l + 2^{2l} + \cdots + 2^{(l'-1)l}) = 2^{ll'} - 1 \equiv 1 \pmod{2^k - 1} .$$

Therefore, $x^{2^{l'l}} = x^2$ for any $x \in \mathrm{GF}(2^k)$ and this identity will be used repeatedly further in the proofs.

In the following lemmas, we always assume that $l < k$ is a positive integer with $\gcd(l,k) = 1$. We also take $A_a(v)$ defined in (5) and $R(v)$ defined in (6). Lemmas 2 and 3 here provide generalization for Lemmas 3, 4 and 6 in [6]. Theorem 1 is a generalization of Lemma 7 in [6].

**Lemma 2** *For any $a \in \mathrm{GF}(2^k)^*$ the element $v_0 = R(a^{-1})$ is a zero of $A_a(v)$ in $\mathrm{GF}(2^k)^*$.*

**Proof.** Since $D(v)$ in (7) is a permutation polynomial on $\mathrm{GF}(2^k)^*$, then for any fixed $a \in \mathrm{GF}(2^k)^*$ the equation

$$av^{2^l+1} = \sum_{i=1}^{l'} v^{2^{il}} + l' + 1 \qquad (8)$$

has exactly one solution $v_0 = R(a^{-1})$ in $\mathrm{GF}(2^k)^*$. Raising (8) to the power of $2^l$ results in

$$a^{2^l} v^{2^{2l}+2^l} = \sum_{i=2}^{l'+1} v^{2^{il}} + l' + 1 = \sum_{i=2}^{l'} v^{2^{il}} + v^{2^{l+1}} + l' + 1 \ .$$

The latter identity, after being added to (8) and setting $v = v_0$, gives

$$av_0^{2^l+1} = a^{2^l} v_0^{2^{2l}+2^l} + v_0^{2^l} + v_0^{2^{l+1}}$$

and consecutively, since $v_0 \neq 0$, $A_a(v_0) = a^{2^l} v_0^{2^{2l}} + v_0^{2^l} + av_0 + 1 = 0$. $\square$

**Lemma 3** *For any $a \in \mathrm{GF}(2^k)^*$ let $z$ be a zero of $A_a(v)$ in $\mathrm{GF}(2^k)$. Then*

$$\mathrm{Tr}_k(z) = \mathrm{Tr}_k(v_0)$$

*and*

$$\begin{aligned}\mathrm{Tr}_k(az^{2^l+1}) &= l'\mathrm{Tr}_k(v_0) + \mathrm{Tr}_k(l'+1) && \text{if} \quad z = v_0 \ , \\ &= l'\mathrm{Tr}_k(v_0) + \mathrm{Tr}_k(l') && \text{if} \quad z \neq v_0 \ ,\end{aligned}$$

*where $v_0 = R(a^{-1})$.*

**Proof.** The first identity follows by observing that any zero of $A_a(v)$ is obtained as a sum of the zero $v_0$ of $A_a(v)$ (see Lemma 2) and a zero of its homogeneous part $a^{2^l} v^{2^{2l}} + v^{2^l} + av$. To prove the identity it therefore suffices to show that $\mathrm{Tr}_k(v_1) = 0$ for any $v_1$ with $a^{2^l} v_1^{2^{2l}} + v_1^{2^l} + av_1 = 0$. This follows from

$$\begin{aligned}\mathrm{Tr}_k(v_1) &= \mathrm{Tr}_k(v_1^{2^{l+1}}) \\ &= \mathrm{Tr}_k(v_1^{2^{2l}+2^l}) \\ &= \mathrm{Tr}_k(a^{2^l} v_1^{2^{2l}+2^l} + av_1^{2^l+1}) \\ &= 0 \ .\end{aligned}$$

To prove the second identity for the case when $z = v_0$ we use the fact presented in the proof of Lemma 2 that $av_0^{2^l+1} = \sum_{i=1}^{l'} v_0^{2^{il}} + l' + 1$. Then $\mathrm{Tr}_k(av_0^{2^l+1}) = l'\mathrm{Tr}_k(v_0) + \mathrm{Tr}_k(l'+1)$.

Now note that since $A_a(v)$ is obtained by adding the $2^l$-th power of (8) to itself we have for $z \neq 0$

$$A_a(z) = 0 \quad \text{if and only if} \quad az^{2^l+1} + \sum_{i=1}^{l'} z^{2^{il}} + l' + 1 \in \{0, 1\}\ .$$

Since $v_0$ is the only solution of (8), then for $z \neq v_0$ with $A_a(z) = 0$ we have $az^{2^l+1} + \sum_{i=1}^{l'} z^{2^{il}} + l' + 1 = 1$ and

$$\mathrm{Tr}_k(az^{2^l+1}) = l'\mathrm{Tr}_k(z) + \mathrm{Tr}_k(l') = l'\mathrm{Tr}_k(v_0) + \mathrm{Tr}_k(l')$$

using already proved identity that $\mathrm{Tr}_k(z) = \mathrm{Tr}_k(v_0)$. $\square$

Now we introduce a particular sequence of polynomials over $\mathrm{GF}(2^k)$ and prove some important properties of these that will be used further for getting the main result of this section about zeros of $A_a(v)$. Denote

$$e(i) = 1 + 2^l + 2^{2l} + \cdots + 2^{(i-1)l} \quad \text{for} \quad i = 1, \ldots, l'$$

so, in particular, $e(l') = (2^l - 1)^{-1} \pmod{2^k - 1}$. Now take every additive term $v^e$ with $e \neq 0$ in the polynomial $1 + (1+v)^{e(i)}$ and replace the exponent $e$ with the cyclotomic equivalent number obtained by shifting the binary expansion of $e$ maximally (till you get an odd number) in the direction of the least significant bits. We call this *reduction* procedure. Recall that two exponents $e_1$ and $e_2$ are cyclotomic equivalent if $2^i e_1 \equiv e_2 \pmod{2^k - 1}$ for some $i < k$. For instance, $v^{2^{il}}$ is reduced to $v$ and $v^{2^{il}+2^{jl}}$ is reduced to $v^{1+2^{(j-i)l}}$ if $i < j$ and so on. The obtained reduced polynomials are denoted as $H_i(v)$ and we use square brackets to denote application of the described reduction procedure to a polynomial, so $H_i(v) = [1 + (1+v)^{e(i)}]$ for $i = 1, \ldots, l'$. The first few polynomials in the sequence (after eliminating all pairs of equal terms) are

$H_1(v) = v$

$H_2(v) = [v + v^{2^l} + v^{1+2^l}] = v + v + v^{1+2^l} = v^{1+2^l}$

$H_3(v) = [v + v^{2^l} + v^{2^{2l}} + v^{1+2^l} + v^{1+2^{2l}} + v^{2^l+2^{2l}} + v^{1+2^l+2^{2l}}]$
$= v + v + v + v^{1+2^l} + v^{1+2^{2l}} + v^{1+2^l} + v^{1+2^l+2^{2l}} = v + v^{1+2^{2l}} + v^{1+2^l+2^{2l}}\ .$

**Lemma 4** *If polynomials $H_i(v)$ are defined as above then*

$$\mathrm{Tr}_k(H_i(v)) = \mathrm{Tr}_k\bigl(1 + (1+v)^{e(i)}\bigr)$$

*for any $v \in \mathrm{GF}(2^k)$ and $i = 1,\ldots,l'$. Also let $Q(v) = (x_0^{2^l+1} + x_0)v^{2^l} + x_0^2 v + x_0$ for any $x_0 \in \mathrm{GF}(2^k)^*$. Then*

$$Q(H_{l'}(x_0^{-1})) = (1 + x_0)(1 + x_0^{-1})^{e(l')} \ .$$

**Proof.** The trace identity for $H_{l'}(v)$ we get obviously from the definition. Further, for any $i \in \{2,\ldots,l'\}$

$$\begin{aligned}
H_i(v) &= [1 + (1+v)^{e(i)}] \\
&= [1 + (1+v)^{e(i-1)}(1+v)^{2^{(i-1)l}}] \\
&= [H_{i-1}(v) + v^{2^{(i-1)l}}(1+v)^{e(i-1)}] \\
&\stackrel{(*)}{=} v(1+v)^{e(i)-1} + H_{i-1}(v) \ ,
\end{aligned}$$

where (*) follows from the following argumentation. First, note that the exponents of additive terms in $v(1+v)^{e(i)-1}$ are exactly all $2^{i-1}$ distinct integers of the form $1 + t_1 2^l + \cdots + t_{i-1} 2^{(i-1)l}$ with $t_j \in \{0,1\}$ for $j = 1,\ldots,i-1$ and the reduction does not apply to any of these so

$$[v(1+v)^{e(i)-1}] = v(1+v)^{e(i)-1} \ .$$

On the other hand, the number of terms in $[v^{2^{(i-1)l}}(1+v)^{e(i-1)}]$ is also equal to $2^{i-1}$ since the exponents in these terms are exactly all the integers of the form $t_0 + t_1 2^l + \cdots + t_{i-2} 2^{(i-2)l} + 2^{(i-1)l}$ with $t_j \in \{0,1\}$ for $j = 0,\ldots,i-2$ and none of these become equal after the reduction. Moreover, every such an exponent, after reduction, can be found in $v(1+v)^{e(i)-1}$ so

$$[v^{2^{(i-1)l}}(1+v)^{e(i-1)}] = v(1+v)^{e(i)-1} \ .$$

Also note that all terms of $H_{i-1}(v)$ are also present in $v(1+v)^{e(i)-1}$. Thus, the number of terms in $H_i(v)$ that remain after eliminating all pairs of equal terms and denoted as $\#H_i$ is equal to $2^{i-1} - \#H_{i-1}$. Unfolding the obtained recursive expression for $H_i(v)$ starting from $H_1(v) = v$ we get that

$$H_i(v) = v(1 + (1+v)^{2^l} + (1+v)^{2^l + 2^{2l}} + \cdots + (1+v)^{e(i)-1}) \ .$$

Now we can evaluate

$$Q(H_{l'}(x_0^{-1})) =$$
$$= (x_0^{2^l+1} + x_0)H_{l'}(x_0^{-1})^{2^l} + x_0^2 H_{l'}(x_0^{-1}) + x_0$$
$$= (x_0 + x_0^{-2^l+1})\left(1 + (1+x_0^{-1})^{2^{2l}} + (1+x_0^{-1})^{2^{2l}+2^{3l}} + \cdots + (1+x_0^{-1})^{2^{2l}+\cdots+2^{l'l}}\right)$$
$$+ x_0\left(1 + (1+x_0^{-1})^{2^l} + (1+x_0^{-1})^{2^l+2^{2l}} + \cdots + (1+x_0^{-1})^{e(l')-1}\right) + x_0$$
$$= \left((x_0 + x_0^{-2^l+1}) + x_0(1+x_0^{-1})^{2^l}\right)\left(1 + (1+x_0^{-1})^{2^{2l}} + \cdots + (1+x_0^{-1})^{2^{2l}+\cdots+2^{(l'-1)l}}\right)$$
$$+ (x_0 + x_0^{-2^l+1})(1+x_0^{-1})^{2^{2l}+\cdots+2^{l'l}} + x_0 + x_0$$
$$= x_0(1+x_0^{-1})^{2^l+2^{2l}+\cdots+2^{l'l}}$$
$$= x_0(1+x_0^{-1})^{2+2^l+2^{2l}+\cdots+2^{(l'-1)l}}$$
$$= (1+x_0)(1+x_0^{-1})^{e(l')}$$

as claimed. □

**Lemma 5** *For any $a \in \mathrm{GF}(2^k)^*$ let $x_0 \in \mathrm{GF}(2^k)$ satisfy $x_0^{2^l+1} + x_0 = a$. Then*

$$\mathrm{Tr}_k\bigl(1 + (1+x_0^{-1})^{e(l')}\bigr) = \mathrm{Tr}_k(R(a^{-1})) \ .$$

**Proof.** Denote $\Gamma = x_0^{2^l-1} + x_0^{-1}$ (obviously $\Gamma \neq 0$ since $x_0 \neq 1$), $\Delta = \Gamma^{-e(l')}$ and further, using Lemma 4, evaluate

$$Q(H_{l'}(x_0^{-1}))x_0^{e(l')} = (1+x_0)(1+x_0)^{e(l')} = (1+x_0^{2^l})^{e(l')}$$

and thus, $Q(H_{l'}(x_0^{-1}))^{2^l-1} = \Gamma$ or, equivalently,

$$Q(H_{l'}(x_0^{-1})) = \Delta^{-1} \ . \tag{9}$$

In what follows, we use the technique suggested by Dobbertin for proving [7, Theorem 1]. Note that

$$A_a(v) = a^{2^l}v^{2^{2l}} + x_0^{2^l+1}v^{2^l} + x_0^{2^l} + (x_0^{2^l-1} + x_0^{-1})\left((x_0^{2^l+1} + x_0)v^{2^l} + x_0^2 v + x_0\right)$$
$$= Q(v)^{2^l} + \Gamma Q(v) = Q(v)(Q(v)^{2^l-1} + \Delta^{-(2^l-1)})$$

for $x_0^{2^l+1} + x_0 = a$ and therefore, by (9), $A_a(H_{l'}(x_0^{-1})) = 0$. Consider the equation

$$Q(v) + \Delta^{-1} = 0 \tag{10}$$

whose roots are also the zeros of $A_a(v)$. We will show that (10) has exactly two roots with $H_{l'}(x_0^{-1})$ and $R(a^{-1})$ being among them (however, we do not claim that $R(a^{-1}) \neq H_{l'}(x_0^{-1})$). Multiplying (10) by $\mu = (x_0^2 \Delta)^{-1}$ and using that $(x_0^{2^l+1} + x_0)\Delta^{2^l-1} = x_0^2$ gives

$$\mu((x_0^{2^l+1} + x_0)v^{2^l} + x_0^2 v + x_0 + \Delta^{-1}) = (v/\Delta)^{2^l} + v/\Delta + x_0\mu + x_0^2\mu^2 = 0 ,$$

which has exactly two solutions $z_0 = H_{l'}(x_0^{-1})$ (see (9)) and $z_1 = H_{l'}(x_0^{-1}) + \Delta$ since its linearized homogeneous part $(v/\Delta)^{2^l} + v/\Delta$ has exactly two roots $v = 0$ and $v = \Delta$. Thus, $z_0 + z_1 = \Delta = \left(\frac{x_0}{1+x_0^{2^l}}\right)^{e(l')}$. Using $(x_0^{2^l} + 1)\Delta^{2^l-1} = x_0$ it is easy to see that $\Delta^{2^l} = x_0\Delta + (x_0\Delta)^{2^l}$ and we have $\mathrm{Tr}_k(\Delta) = 0$.

Now we show that none of the possible roots of $Q(v) = 0$ is a solution of (8). In fact, suppose that $Q(z) = 0$. Then, since $x_0 \neq 0$, we have $z^{2^l} = (x_0 z)^{2^l} + x_0 z + 1$ and $az^{2^l} = x_0^2 z + x_0$ (since $a = x_0^{2^l+1} + x_0$). We put such a $z$ into (8) and compute

$$az^{2^l+1} + \sum_{i=1}^{l'} z^{2^{il}} + l' + 1$$

$$= (x_0^2 z + x_0)z + \sum_{i=0}^{l'-1}(x_0 z)^{2^{il}} + \sum_{i=1}^{l'}(x_0 z)^{2^{il}} + l' + l' + 1$$

$$= 1 .$$

Therefore, recalling the proved identity $A_a(v) = Q(v)(Q(v)^{2^l-1} + \Delta^{-(2^l-1)})$ and keeping in mind that $\gcd(2^l - 1, 2^k - 1) = 1$ we see that $v_0 = R(a^{-1})$ which is the unique solution of (8) and, by Lemma 2, also the root of $A_a(v) = 0$, satisfies $Q(v_0) = \Delta^{-1}$. Recall that (10) has exactly two solutions $z_0 = H_{l'}(x_0^{-1})$ and $z_1 = H_{l'}(x_0^{-1}) + \Delta$. Thus, $R(a^{-1}) + H_{l'}(x_0^{-1}) = \Delta$ or $R(a^{-1}) = H_{l'}(x_0^{-1})$ (although we do not need in our proof that $R(a^{-1}) \neq H_{l'}(x_0^{-1})$, we believe that this holds) and, by Lemma 4,

$$\mathrm{Tr}_k(R(a^{-1})) = \mathrm{Tr}_k(H_{l'}(x_0^{-1})) = \mathrm{Tr}_k(1 + (1 + x_0^{-1})^{e(l')})$$

as claimed. □

**Theorem 1** *For any $a \in \mathrm{GF}(2^k)^*$ and a positive integer $l < k$ with $\gcd(l, k) = 1$, let $A_a(v)$ be defined as in (5). Also let*

$$M_i = \{a \mid A_a(v) \text{ has exactly } i \text{ zeros in } \mathrm{GF}(2^k)\} . \tag{11}$$

*Then $A_a(v)$ has either one, two or four zeros in $\mathrm{GF}(2^k)$. For $i \in \{1, 2, 4\}$, we have $a \in M_i$ if and only if $p_a(x) = x^{2^l+1} + x + a$ has exactly $i - 1$ zeros in $\mathrm{GF}(2^k)$. The following distribution holds for $k$ odd (resp. $k$ even)*

$$\begin{aligned}
|M_1| &= \tfrac{2^k+1}{3} & (\text{resp. } \tfrac{2^k-1}{3}) , \\
|M_2| &= 2^{k-1} - 1 & (\text{resp. } 2^{k-1}) , \\
|M_4| &= \tfrac{2^{k-1}-1}{3} & (\text{resp. } \tfrac{2^{k-1}-2}{3}) .
\end{aligned}$$

*Furthermore, $a \in M_2$ if and only if $\mathrm{Tr}_k(R(a^{-1}) + 1) = 1$, where $R(v)$ is defined in (6).*

**Proof.** In Lemma 2 it was shown that $v_0 = R(a^{-1})$ is a zero of $A_a(v)$ in $\mathrm{GF}(2^k)^*$. Let $N_a$ be the number of zeros of $A_a(v)$ in $\mathrm{GF}(2^k)$. Since $A_a(v)$ has a zero in $\mathrm{GF}(2^k)$, $N_a$ is equal to the number of zeros of its homogeneous part $a^{2^l}v^{2^{2l}} + v^{2^l} + av$ in $\mathrm{GF}(2^k)$. Dividing the latter polynomial by $a^{-1}v$, then raising it to power $2^{k-1}$ and replacing $(av^{2^l-1})^{2^{k-1}}$ by $x$ leads to

$$p_a(x) = x^{2^l+1} + x + a \; ,$$

which, since $\gcd(2^l - 1, 2^k - 1) = 1$, has $N_a - 1$ zeros in $\mathrm{GF}(2^k)$. It is therefore sufficient to study the number of zeros of this polynomial in $\mathrm{GF}(2^k)$.

From now on assume that $N_a \geq 2$. Then $p_a(x)$ has a zero $x_0 \in \mathrm{GF}(2^k)$. Now we replace $x$ in $p_a(x)$ with $x + x_0$ to get

$$(x + x_0)^{2^l+1} + (x + x_0) + a = 0$$

or

$$x^{2^l+1} + x_0 x^{2^l} + x_0^{2^l} x + x_0^{2^l+1} + x + x_0 + a = 0$$

which implies

$$x^{2^l+1} + x_0 x^{2^l} + (x_0^{2^l} + 1)x = 0 \; .$$

Since $x = 0$ corresponds to $x_0$ being the zero of $p_a(x)$, we can divide the latter equation by $x$ and after substituting $y = x^{-1}$ we note that if $p_a(x)$ has a zero then the reciprocal equation, given by

$$(x_0^{2^l} + 1)y^{2^l} + x_0 y + 1 = 0 \tag{12}$$

has $N_a - 2$ zeros. This affine equation has either zero roots in $\mathrm{GF}(2^k)$ or the same number of roots as its homogeneous part $(x_0^{2^l} + 1)y^{2^l} + x_0 y$ which is seen to have exactly two solutions, the zero solution and a unique nonzero solution, since $\gcd(2^l - 1, 2^k - 1) = 1$. Therefore, it can be concluded that $p_a(x) = 0$ can have either zero, one or three solutions or, equivalently, $A_a(v)$ has either one, two or four zeros in $\mathrm{GF}(2^k)$.

Now we need to find the conditions when there exists a solution of (12). Let $y = tw$, where $t^{2^l-1} = c$ and $c = \frac{x_0}{x_0^{2^l}+1}$. Since $\gcd(2^l - 1, 2^k - 1) = 1$, there is a one-to-one correspondence between $t$ and $c$. Then (12) is equivalent to

$$w^{2^l} + w + \frac{1}{ct(x_0^{2^l}+1)} = 0 \; .$$

Hence, (12) has no solutions if and only if

$$\text{Tr}_k\left(\frac{1}{ct(x_0^{2^l}+1)}\right) = 1 .$$

This easily follows from the fact that the linear operator $L(\omega) = \omega^{2^l} + \omega$ on $\text{GF}(2^k)$ has the kernel of dimension one and, thus, the number of elements in the image of $L$ is $2^{k-1}$. Since all the elements $\omega^{2^l} + \omega$ have the trace zero and the total number of such elements in $\text{GF}(2^k)$ is $2^{k-1}$, we conclude that the image of $L$ contains all the elements in $\text{GF}(2^k)$ having trace zero.

Since $c = t^{2^l-1}$ then $t = c^{e(l')}$. Thus, from the definition of $c$ and $t$ we get

$$\text{Tr}_k\left(\frac{1}{ct(x_0^{2^l}+1)}\right) = \text{Tr}_k\left(\left(\frac{x_0^{2^l}+1}{x_0}\right)^{1+e(l')}\left(\frac{1}{x_0^{2^l}+1}\right)\right)$$

$$= \text{Tr}_k\left(\frac{(x_0^{2^l}+1)^{e(l')}}{x_0^{1+e(l')}}\right) = \text{Tr}_k\left(\frac{(x_0+1)^{2^l e(l')}}{x_0^{2^l e(l')}}\right) = \text{Tr}_k\left((1+x_0^{-1})^{e(l')}\right) .$$

We conclude that $p_a(x)$ has exactly one zero (which is $x_0$) if and only if

$$\text{Tr}_k\left((1+x_0^{-1})^{e(l')}\right) = 1 . \tag{13}$$

It means that $A_a(v)$ has exactly two zeros in $\text{GF}(2^k)$ (i.e., $N_a = 2$) only for such $a$ that $a = x_0^{2^l+1} + x_0$ with (13) holding. Combining this with the result of Lemma 5, we conclude that $A_a(v)$ has exactly two zeros in $\text{GF}(2^k)$ if and only if

$$\text{Tr}_k(R(a^{-1})+1) = 1 .$$

In the case of one or four zeros, $\text{Tr}_k(R(a^{-1})+1) = 0$.

Now note that since $e(l') = 1+2^l+2^{2l}+\cdots+2^{(l'-1)l}$ is invertible modulo $2^k-1$ with the multiplicative inverse equal to $2^l-1$ then $\gcd(e(l'), 2^k-1) = 1$ and thus, $(1+v^{-1})^{e(l')}$ is a one-to-one mapping of $\text{GF}(2^k)^*$ onto $\text{GF}(2^k)\setminus\{1\}$. Therefore, if $k$ is odd (resp. $k$ is even) then the number of $x_0 \in \text{GF}(2^k)^*$ satisfying (13) is equal to $2^{k-1}-1$ (resp. $2^{k-1}$) and obviously $x_0 \neq 1$. On the other hand, if $N_a = 2$ then $x^{2^l+1} + x = a$ has a unique solution $x_0$ and so the number of nonzero values $a \in \text{GF}(2^k)^*$ with $N_a = 2$ for $k$ odd (resp. $k$ even) is $|M_2| = 2^{k-1}-1$ (resp. $2^{k-1}$). Now note that if $a = 0$ then $p_a(x) = x^{2^l+1} + x + a$ has exactly two zeros $x = \{0, 1\}$. Thus, considering the mapping $x \mapsto x^{2^l+1} + x$ for $x$ running through $\text{GF}(2^k)\setminus\{0,1\}$ it is easy to see that $|M_2| + 3|M_4| = 2^k - 2$ and, knowing $|M_2|$, we can find $|M_4|$. Finally, the last remaining unknown $|M_1|$ can be evaluated from the obvious equation $|M_1| + |M_2| + |M_4| = |\text{GF}(2^k)^*| = 2^k - 1$. □

Note the paper [9] by Bluher where $x^{p^l+1} + ax + b$ and the related polynomials similar to the linearized part of $A_a(v)$ over an arbitrary field of characteristic $p$ are studied. In particular, the possible number of zeros and corresponding values of $|M_i|$, in the notations of our Theorem 1, were found (see [9, Theorems 5.6, 6.4]). This was also done earlier for odd $k$ in [10, Lemma 9].

# 4 The Linearized Polynomial $L_a(z)$

The distribution of the three-valued crosscorrelation function to be determined in Section 5 depends on the detailed distribution of the number of zeros in $\mathrm{GF}(2^m)$ of the linearized polynomial

$$L_a(z) = z^{2^{k+l}} + r^{2^l} a^{2^l} z^{2^{2l}} + raz , \qquad (14)$$

where $a \in \mathrm{GF}(2^k)$, $r \in \mathrm{GF}(2^m)$ and $m = 2k$. Some additional conditions on the parameters will be imposed later. For the details on linearized polynomials in general, the reader is referred to Lidl and Niederreiter [11]. In the following lemmas, we always take $L_a(z)$ defined in (14).

**Lemma 6** *Let $l$ and $k$ be integers with $\gcd(l, k) = 1$, $a \in \mathrm{GF}(2^k)$ and $r \in \mathrm{GF}(2^m)$. If $L_a(z) = 0$ for some $z \in \mathrm{GF}(2^m)$ then*

$$a\mathrm{Tr}_k^m(rz^{2^l+1}) \in \{0, 1\} ,$$

*where $\mathrm{Tr}_k^m(x) = x + x^{2^k}$ is a trace mapping from $\mathrm{GF}(2^m)$ to $\mathrm{GF}(2^k)$.*

**Proof.** For any $z \in \mathrm{GF}(2^m)$ with $L_a(z) = 0$ we have

$$z^{2^l} L_a(z) = raz^{2^l+1} + (raz^{2^l+1})^{2^l} + z^{2^l(2^k+1)} = 0$$

and $z^{2^l(2^k+1)} \in \mathrm{GF}(2^k)$. Thus, $\mathrm{Tr}_k^m(raz^{2^l+1}) + (\mathrm{Tr}_k^m(raz^{2^l+1}))^{2^l} = 0$ meaning that $a\mathrm{Tr}_k^m(rz^{2^l+1}) \in \mathrm{GF}(2^l) \cap \mathrm{GF}(2^k) = \{0, 1\}$. □

**Lemma 7** *Let $l$ and $k$ be odd with $\gcd(l, k) = 1$, $a \in \mathrm{GF}(2^k)$ and $r$ be a noncube in $\mathrm{GF}(2^m)$ such that $r^{2^k+1} = 1$. Then the following holds.*

(i) *The number of zeros of $L_a(z)$ in $\mathrm{GF}(2^m)$ is 1 or 4.*

(ii) *If, additionally, $a \neq 0$ and $\mathrm{Tr}_k(v_0) = 0$ (where $v_0 = R(a^{-1})$ and $R(v)$ is defined in (6)) then $L_a(z)$ has $z = 0$ as its only zero in $\mathrm{GF}(2^m)$.*

**Proof.** First of all, let $\bar{z} = z^{2^k}$ for any $z \in \mathrm{GF}(2^m)$ and also let $U = rz^{2^l+1}$. If $z \neq 0$ and $L_a(z) = 0$ then, since $l$ is odd and $r$ is a noncube in $\mathrm{GF}(2^m)$ with $r^{2^k+1} = 1$ we have that $U \neq \bar{U}$ and thus, by Lemma 6, and denoting $V = aU$

$$a\mathrm{Tr}_k^m(U) = V + V^{2^k} = 1 . \qquad (15)$$

(i) If $a = 0$ then $L_a(z)$ has a unique zero root so we further assume that $a \neq 0$. The polynomial $L_a(z)$ is a linearized polynomial and its zeros form a vector subspace over $\mathrm{GF}(2)$ (and even over $\mathrm{GF}(2^2)$ since $k + l$ is even). We will study the number of solutions of $L_a(z) = 0$ in $\mathrm{GF}(2^m)$. Note that $L_a(z) = 0$ is equivalent to
$$\overline{z}^{2^l} = r^{2^l} a^{2^l} z^{2^{2l}} + raz .$$

Further, we obtain

$$\begin{aligned}
\overline{U}^{2^l} &= r^{-2^l} \overline{z}^{2^l(2^l+1)} \\
&= r^{-2^l} (r^{2^l} a^{2^l} z^{2^{2l}} + raz)^{2^l+1} \\
&= r^{-2^l} (r^{2^l(2^l+1)} a^{2^l(2^l+1)} z^{2^{2l}(2^l+1)} + r^{2^{2l}+1} a^{2^{2l}+1} z^{2^{3l}+1}) \\
&\quad + r^{-2^l} (r^{2^{l+1}} a^{2^{l+1}} z^{2^{2l}+2^l} + r^{2^l+1} a^{2^l+1} z^{2^l+1}) \\
&= r^{2^{2l}} a^{2^l(2^l+1)} z^{2^{2l}(2^l+1)} + r^{2^{2l}-2^l+1} a^{2^{2l}+1} z^{2^{3l}+1} + r^{2^l} a^{2^{l+1}} z^{2^{2l}+2^l} + r a^{2^l+1} z^{2^l+1} \\
&= a^{2^l(2^l+1)} U^{2^{2l}} + a^{2^{2l}+1} U^{2^{2l}-2^l+1} + a^{2^{l+1}} U^{2^l} + a^{2^l+1} U .
\end{aligned}$$

From now on assume $z \neq 0$. Since $a^{-2^l} = U^{2^l} + \overline{U}^{2^l}$ we have

$$\begin{aligned}
1 &= a^{2^l} U^{2^l} + a^{2^l} \overline{U}^{2^l} \\
&= a^{2^l} U^{2^l} + a^{2^{2l}+2^{l+1}} U^{2^{2l}} + a^{2^{2l}+2^l+1} U^{2^{2l}-2^l+1} + a^{2^{l+1}+2^l} U^{2^l} + a^{2^{l+1}+1} U
\end{aligned}$$

which leads to

$$a^{2^{2l}+2^{l+1}} U^{2^{2l}} + a^{2^{2l}+2^l+1} U^{2^{2l}-2^l+1} + (a^{2^l} + a^{2^{l+1}+2^l}) U^{2^l} + a^{2^{l+1}+1} U + 1 = 0 .$$

Substituting $V = aU$ and multiplying by $b = a^{-2^{l+1}}$, simplifies the equation to

$$V^{2^{2l}} + V^{2^{2l}-2^l+1} + (1+b) V^{2^l} + V + b = 0$$

which after multiplying by $V^{2^l}$ gives

$$(V^{2^l} + V)^{2^l+1} + bV^{2^l}(V^{2^l} + 1) = 0 . \qquad (16)$$

Since

$$\frac{(V^{2^l} + V)^{2^l+1}}{V^{2^l}(V^{2^l} + 1)} = (V+1)^{2^{2l}-2^l+1} + V^{2^{2l}-2^l+1} + 1$$

we obtain

$$(V+1)^{2^{2l}-2^l+1} + V^{2^{2l}-2^l+1} + 1 = b .$$

As proved in [7, Corollary 2], the monomial function $f(x) = x^{2^{2l}-2^l+1}$ is almost perfect nonlinear (APN) when $\gcd(l, m) = 1$, which is the case here since $l$ is odd and $\gcd(l, k) = 1$. This means that the number of solutions $V \in \mathrm{GF}(2^m)$ of the latter equation is at most 2 for any $b$ in $\mathrm{GF}(2^m)$. Since $V = raz^{2^l+1}$ and $\gcd(2^l + 1, 2^m - 1) = 3$ it follows that the number of zeros in $\mathrm{GF}(2^m)^*$ of the linearized polynomial $L_a(z)$ is at most 6, which implies that the number of zeros in $\mathrm{GF}(2^m)$ is 1 or 4 since the zeros of $L_a(z)$ form a vector subspace over $\mathrm{GF}(2^2)$.

(ii) Let $x = V^{2^l} + V$ and assume $z \neq 0$. After rewriting $z^{2^l} L_a(z) = z^{2^l(2^k+1)} + x = 0$ observe that this implies that $x \in \mathrm{GF}(2^k)$.

Using (16) we have

$$b^{-1} x^{2^l+1} = \sum_{i=1}^{\tilde{l}} x^{2^{il}} \;,$$

where $\tilde{l} l = 1 (\bmod\ m)$. Such an $\tilde{l}$ exists since $l$ is odd, $\gcd(l, k) = 1$ and therefore, $\gcd(l, m) = 1$. Raising the latter identity to the power $2^l$ and adding to itself implies

$$c^{2^l} x^{2^{2l}+2^l} + c x^{2^l+1} = x^{2^{(\tilde{l}+1)l}} + x^{2^l} \;,$$

where $c = b^{-1} = a^{2^l+1}$. Dividing by $x^{2^l}$ ($x \neq 0$ since otherwise the only zero of $L_a(z)$ is $z = 0$) implies

$$c^{2^l} x^{2^{2l}} + x^{2^l} + cx + 1 = 0 \;.$$

By Theorem 1, the latter equation has exactly two roots in $\mathrm{GF}(2^k)$ if and only if $\mathrm{Tr}_k(R(c^{-1})) = \mathrm{Tr}_k(R(a^{-2^{l+1}})) = \mathrm{Tr}_k(v_0) = 0$ and $R(a^{-2^{l+1}}) = v_0^{2^{l+1}}$ is one of its roots. From Lemma 3 it also follows that all the roots of this equation have the same trace as $v_0$. Therefore, in the case when $\mathrm{Tr}_k(v_0) = 0$ both roots have trace zero. However, since $x = V^{2^l} + V \in \mathrm{GF}(2^k)$ and $V \notin \mathrm{GF}(2^k)$ (recall that $V = aU$ and $U \neq \overline{U}$) we have

$$\mathrm{Tr}_k(x) = \sum_{i=0}^{k-1} x^{2^i} = \sum_{i=0}^{k-1} x^{2^{li}} = \sum_{i=0}^{k-1} (V^{2^{l(i+1)}} + V^{2^{li}})$$

$$= V^{2^{kl}} + V \stackrel{(*)}{=} 1 \neq \mathrm{Tr}_k(v_0) \;,$$

where (*) holds since $V^{2^{lk}} = V^{2^k}$ for odd $l$ and $V^{2^k} + V \neq 0$ if $V \notin \mathrm{GF}(2^k)$. Therefore, if $\mathrm{Tr}_k(v_0) = 0$ then there is no solutions $x \in \mathrm{GF}(2^k)$ having the form $x = V^{2^l} + V$. We have therefore shown that in the case $\mathrm{Tr}_k(v_0) = 0$ there are no nonzero solutions of $L_a(z) = 0$ in $\mathrm{GF}(2^m)$. □

## 5 Three-Valued Crosscorrelation

In this section, we prove our main result formulated in Corollary 1. We start by considering the following exponential sum denoted $S_0(a)$ that to some extent is determined by the following lemma that repeats Lemma 10 in [6]. It is assumed everywhere that $m = 2k$.

**Lemma 8 ([6])** *For an odd $k$, integer $l < k$ and $a \in \mathrm{GF}(2^k)$ let $S_0(a)$ be defined by*

$$S_0(a) = \sum_{y \in \mathrm{GF}(2^m)} (-1)^{\mathrm{Tr}_m(ay^{2^l+1}) + \mathrm{Tr}_k(y^{2^k+1})} .$$

*Then*

$$S_0(a) = 2^k \sum_{v \in \mathrm{GF}(2^k), A_a(v)=0} (-1)^{\mathrm{Tr}_k(a(l+1)v^{2^l+1}+v)} ,$$

*where $A_a(v)$ is defined in (5).*

We can now determine $S_0(a)$ completely in the following corollary.

**Corollary 2** *Under the conditions of Lemma 8 and, additionally, assuming $a \neq 0$ and $\gcd(l, k) = 1$ let $M_i$ be defined as in (11). Then the distribution of $S_0(a)$ for $l$ even is as follows:*

$$\begin{array}{ll} -2^{k+1} & \text{if } a \in M_4 , \\ 0 & \text{if } a \in M_2 , \\ 2^k & \text{if } a \in M_1 \end{array}$$

*and for $l$ odd*

$$\begin{array}{ll} -2^{k+2} & \text{if } a \in M_4 , \\ 2^{k+1} & \text{if } a \in M_2 , \\ -2^k & \text{if } a \in M_1 . \end{array}$$

**Proof.** Let $l' = l^{-1} \pmod{k}$. The distribution follows directly from Lemmas 3 and 8 since these imply that for $l$ even

$$S_0(a) = 2^k (-1)^{(l'+1)\mathrm{Tr}_k(v_0) + l'} (N_a - 2)$$

and for $l$ odd

$$S_0(a) = 2^k (-1)^{\mathrm{Tr}_k(v_0)} N_a ,$$

where $N_a$ is the number of zeros of $A_a(v)$ in $\mathrm{GF}(2^k)$ and $v_0 = R(a^{-1})$. Finally, using Theorem 1, we get the claimed result. □

**Lemma 9** *Let $k$ be odd and $r$ be a noncube in $\mathrm{GF}(2^m)$ such that $r^{2^k+1} = 1$. Let also $a \in \mathrm{GF}(2^k)$ and*

$$S_1(a) = \sum_{y \in \mathrm{GF}(2^m)} (-1)^{\mathrm{Tr}_m(ray^{2^l+1}) + \mathrm{Tr}_k(y^{2^k+1})} ,$$

$$S_2(a) = \sum_{y \in \mathrm{GF}(2^m)} (-1)^{\mathrm{Tr}_m(r^{-1}ay^{2^l+1}) + \mathrm{Tr}_k(y^{2^k+1})} .$$

*Then*

(i) $S_1(a) = S_2(a)$.

(ii) Furthermore, if, additionally, $l$ is odd with $\gcd(l,k) = 1$ then for $i = 1, 2$ holds
$$S_i(a)^2 = 2^m T_a\ ,$$
where $T_a$ is the number of zeros in $\mathrm{GF}(2^m)$ of $L_a(z)$ defined in (14).

**Proof.** (i) Using definitions, straightforward calculations lead to
$$S_1(a) = \sum_{y \in \mathrm{GF}(2^m)} (-1)^{\mathrm{Tr}_m(ray^{2^l}+1) + \mathrm{Tr}_k(y^{2^k}+1)}$$
$$= \sum_{y \in \mathrm{GF}(2^m)} (-1)^{\mathrm{Tr}_m(r^{2^k} a^{2^k} y^{(2^l+1)2^k}) + \mathrm{Tr}_k(y^{(2^k+1)2^k})}$$
$$= \sum_{z \in \mathrm{GF}(2^m)} (-1)^{\mathrm{Tr}_m(r^{-1}az^{2^l}+1) + \mathrm{Tr}_k(z^{2^k}+1)}$$
$$= S_2(a)\ .$$

(ii) First, it can be noticed that here we are with the hypothesis of Lemma 7 Item (i). Using substitution $z = x + y$ we obtain
$$S_1(a)^2 = \sum_{x,y \in \mathrm{GF}(2^m)} (-1)^{\mathrm{Tr}_m(ra(x^{2^l}+1+y^{2^l}+1)) + \mathrm{Tr}_k(x^{2^k}+1+y^{2^k}+1)}$$
$$= \sum_{y,z \in \mathrm{GF}(2^m)} (-1)^{\mathrm{Tr}_m(ra((z+y)^{2^l}+1+y^{2^l}+1)) + \mathrm{Tr}_k((z+y)^{2^k}+1+y^{2^k}+1)}$$
$$= \sum_{y,z \in \mathrm{GF}(2^m)} (-1)^{\mathrm{Tr}_m(ra(z^{2^l}y+zy^{2^l}+z^{2^l}+1)+yz^{2^k}) + \mathrm{Tr}_k(z^{2^k}+1)}$$
$$= \sum_{z \in \mathrm{GF}(2^m)} (-1)^{\mathrm{Tr}_m(raz^{2^l}+1) + \mathrm{Tr}_k(z^{2^k}+1)} \sum_{y \in \mathrm{GF}(2^m)} (-1)^{\mathrm{Tr}_m(y^{2^l}(z^{2^{k+l}}+r^{2^l}a^{2^l}z^{2^{2l}}+raz))}$$
$$= 2^m \sum_{z \in \mathrm{GF}(2^m),\, L_a(z)=0} (-1)^{\mathrm{Tr}_m(raz^{2^l}+1) + \mathrm{Tr}_k(z^{2^k}+1)}\ ,$$

where $L_a(z) = z^{2^{k+l}} + r^{2^l}a^{2^l}z^{2^{2l}} + raz$.

It remains to show that $f(z) = \mathrm{Tr}_m(raz^{2^l+1}) + \mathrm{Tr}_k(z^{2^k+1}) = 0$ for any root $z$ of $L_a$. If $z = 0$ then this fact is obvious. If $z \neq 0$ then, by (15) from Lemma 7, $\mathrm{Tr}_k^m(V) = V + V^{2^k} = 1$, where $V = raz^{2^l+1}$ implying that $\mathrm{Tr}_m(V) = 1$. Moreover, multiplying $L_a(z) = 0$ by $z^{2^l}$ we obtain $V + V^{2^l} + z^{2^l(2^k+1)} = 0$. Thus,
$$f(z) = 1 + \mathrm{Tr}_k(z^{2^k+1}) = 1 + \mathrm{Tr}_k(V + V^{2^l})\ .$$

But

$$\mathrm{Tr}_k(V + V^{2^l}) =$$
$$= (V + \cdots + V^{2^l} + \cdots + V^{2^{k-1}}) + (V^{2^l} + \cdots + V^{2^{k-1}} + \cdots + V^{2^{l+k-1}})$$
$$= (V + V^{2^k}) + (V + V^{2^k})^2 + \cdots + (V + V^{2^k})^{2^{l-1}} = l \pmod{2} = 1$$

and thus, $f(z) = 0$.

In particular, since $S_1(a) = \sum_{y \in \mathrm{GF}(2^m)}(-1)^{f(y)} \neq 0$ the Boolean function $f(z)$ can not be balanced. Quadratic functions including those similar to $f(z)$ are studied in [12]. $\square$

We are now in position to completely determine the distribution of $S(a)$ defined in (4) for $a \in \mathrm{GF}(2^k)^*$. Since this is equivalent to the distribution of $C_d(\tau) + 1$ for $\tau = 0, 1, \ldots, 2^k - 2$, our main result in Corollary 1 is a consequence of the theorem below. Note that for any $d$ with the prescribed property we have $\gcd(d, 2^k - 1) = 1$.

**Theorem 2** *Let $m = 2k$ and $d(2^l + 1) \equiv 2^i \pmod{2^k - 1}$ for some odd $k$ and integer $l$ with $0 < l < k$, $\gcd(l, k) = 1$ and $i \geq 0$. Then the exponential sum $S(a)$ defined in (4) for $a \in \mathrm{GF}(2^k)^*$ (and $C_d(\tau) + 1$ for $\tau = 0, 1, \ldots, 2^k - 2$) have the following distribution*

$$\begin{array}{lll} -2^{k+1} & occurs \ \frac{2^{k-1}-1}{3} & times, \\ 0 & occurs \ 2^{k-1} - 1 & times, \\ 2^k & occurs \ \frac{2^k+1}{3} & times. \end{array}$$

**Proof:** To determine the distribution of the crosscorrelation function $C_d(\tau) + 1$ we need to compute the distribution of $S(a)$ as in (4) for $a \in \mathrm{GF}(2^k)^*$. We divide the proof into two cases depending on the parity of $l$.

**Case 1:** ($l$ even)

In this case, $\gcd(2^l + 1, 2^m - 1) = 1$. Therefore, substituting $x = y^{2^l+1}$ in the expression for $S(a)$ and since $d(2^l + 1)(2^k + 1) \equiv 2^i(2^k + 1) \pmod{2^m - 1}$, we are lead to

$$S(a) = \sum_{y \in \mathrm{GF}(2^m)} (-1)^{\mathrm{Tr}_m(ay^{2^l+1}) + \mathrm{Tr}_k(y^{2^k+1})} = S_0(a) ,$$

where $S_0(a)$ is defined in Lemma 8. The distribution of $S(a)$ for even values of $l$ follows, therefore, from the distribution of $S_0(a)$ given in Corollary 2.

**Case 2:** ($l$ odd)

To calculate $S(a)$, we first observe that $\gcd(2^l + 1, 2^m - 1) = 3$. Therefore, if we let $x = y^{2^l+1}$, then $x$ runs through all cubes in $\mathrm{GF}(2^m)$ three times when $y$ runs through $\mathrm{GF}(2^m)$. Thereafter, let $x = ry^{2^l+1}$, where $r$ is a noncube in $\mathrm{GF}(2^m)$ and finally $x = r^{-1}y^{2^l+1}$. When $y$ runs through $\mathrm{GF}(2^m)$ then $x$ will run through GF($2^m$) three times. We select $r$ as a noncube in GF($2^m$) such that $r^{2^k+1} = 1$. Further, since $d(2^l + 1)(2^k + 1) \equiv 2^i(2^k + 1) \pmod{2^m - 1}$, we obtain

$$3S(a) = \sum_{y \in \text{GF}(2^m)} (-1)^{\text{Tr}_m(ay^{2^l+1}) + \text{Tr}_k(y^{2^k+1})}$$

$$+ \sum_{y \in \text{GF}(2^m)} (-1)^{\text{Tr}_m(ray^{2^l+1}) + \text{Tr}_k(y^{2^k+1})}$$

$$+ \sum_{y \in \text{GF}(2^m)} (-1)^{\text{Tr}_m(r^{-1}ay^{2^l+1}) + \text{Tr}_k(y^{2^k+1})}$$

$$= \sum_{i=0}^{2} S_i(a) \; ,$$

where $S_i(a)$ are defined as in Lemmas 8 and 9.

By Lemma 9 we also have that $S_1(a) = S_2(a)$ and

$$S_1(a)^2 = 2^m T_a \; ,$$

where $T_a$ is the number of zeros in GF($2^m$) of $L_a(z) = z^{2^{k+l}} + r^{2^l} a^{2^l} z^{2^{2l}} + raz$. From Lemma 7 Item (i) it follows that $T_a = 1$ or $T_a = 4$ and, therefore, by Lemma 9, we have $S_1(a) = S_2(a) = \pm 2^k$ or $S_1(a) = S_2(a) = \pm 2^{k+1}$.

**Case a:** In the case when $\text{Tr}_k(v_0) = 0$, where $v_0 = R(a^{-1})$ and $R(v)$ is defined in (6), which by Theorem 1, occurs for $2^{k-1} - 1$ distinct values of $a \in M_2$, it follows from Lemma 7 Item (ii) that $T_a = 1$. Therefore, by Lemma 9 we have $S_1^2(a) = 2^m$, i.e., $S_1(a) = S_2(a) = \pm 2^k$. Since $a \in M_2$ and by Corollary 2, $S_0(a) = 2^{k+1}$. Furthermore, since $S(a) = (S_0(a) + S_1(a) + S_2(a))/3$ is an integer, it follows that only $S_1(a) = S_2(a) = -2^k$ is possible and, therefore, $S(a) = 0$.

**Case b:** In the case when $\text{Tr}_k(v_0) = 1$ and $A_a(v) = a^{2^l} v^{2^{2l}} + v^{2^l} + av + 1$ has four zeros in GF($2^k$), which by Theorem 1, occurs for $(2^{k-1} - 1)/3$ distinct values of $a \in M_4$, by Corollary 2 we have $S_0(a) = -2^{k+2}$. Since $S_1(a) = S_2(a) = \pm 2^k$ or $S_1(a) = S_2(a) = \pm 2^{k+1}$ and $S(a)$ is an integer, only two of the four sign combinations are possible, leading in this case to $S(a) = 0$ or $S(a) = -2^{k+1}$.

**Case c:** In the case when $\text{Tr}_k(v_0) = 1$ and $A_a(v)$ has one zero in GF($2^k$), which by Theorem 1, occurs for $(2^k + 1)/3$ distinct values of $a \in M_1$, Corollary 2 gives $S_0(a) = -2^k$. Since $S_1(a) = S_2(a) = \pm 2^k$ or $S_1(a) = S_2(a) = \pm 2^{k+1}$ and $S(a)$ is an integer, only two of the four sign combinations are possible, leading to $S(a) = -2^k$ or $S(a) = 2^k$.

The three cases above give in total the possible values $0, \pm 2^k, -2^{k+1}$ for $S(a)$. We next use the expressions for the sum and the square sum of $C_d(\tau) + 1$ to obtain a set of equations to determine the complete correlation distribution.

Suppose the crosscorrelation function $C_d(\tau) + 1$ takes on the value zero $r$ times, the value $2^k$ is taken on $s$ times, the value $-2^k$ occurs $t$ times and the

Table 1: Exponents $d$ giving three-valued crosscorrelation

| $m$ | Proved in [4] | Proved in [6] | Newly found |
|---|---|---|---|
| 6 | 3 | 3 | |
| 10 | 11 | 7 | |
| 14 | 43 | 15 | 27 |
| 18 | 171 | 31 | 103 |
| 22 | 683 | 63 | 231, 365, 411 |
| 26 | 2731 | 127 | 911, 1243, 1639 |

value $-2^{k+1}$ occurs $v$ times. From Lemma 1 it follows that

$$\begin{aligned} r + s + t + v &= 2^k - 1 \\ 2^k s - 2^k t - 2^{k+1} v &= 2^k \\ 2^{2k} s + 2^{2k} t + 2^{2k+2} v &= 2^m(2^k - 1) \ . \end{aligned}$$

This implies

$$\begin{aligned} r + s + t + v &= 2^k - 1 \\ s - t - 2v &= 1 \\ s + t + 4v &= 2^k - 1 \ . \end{aligned}$$

Since $S(a) = \pm 2^k$ is only possible in Case 3, when $\text{Tr}_k(v_0) = 1$ and $A_a(v)$ has one zero in $\text{GF}(2^k)$, which occurs $(2^k + 1)/3$ times, we get $s + t = (2^k + 1)/3$. From the last equation this leads to $v = (2^{k-1} - 1)/3$ and therefore from the first equation $r = 2^{k-1} - 1$. Finally, using the second equation, we get $t = 0$ and $s = (2^k + 1)/3$. □

In the following, we conjecture that all the cases with the three-valued crosscorrelation fall under the conditions of our main theorem. The conjecture has been verified numerically for all $m \leq 26$ and these results are presented in Table 1.

**Conjecture 1** *Only those cases described in Corollary 1 lead to the three-valued crosscorrelation between two m-sequences of different lengths $2^m - 1$ and $2^k - 1$, where $m = 2k$.*

# 6 Conclusion

We have identified new pairs of $m$-sequences having different lengths $2^m - 1$ and $2^k - 1$, where $m = 2k$, with three-valued crosscorrelation and we have completely determined the crosscorrelation distribution. These pairs differ from the sequences in the Kasami family by the property that instead of the decimation $d = 1$ we take such a $d$ that $d(2^l + 1) \equiv 2^i \pmod{2^k - 1}$ for some integer $l$ and $i \geq 0$, where $k$ is odd and $\gcd(l, k) = 1$. We conjecture that our result covers all the three-valued cases for the crosscorrelation of $m$-sequences with the described parameters.

# Acknowledgment

The authors would like to thank the anonymous reviewers for suggesting a shorter proof of Lemmas 6 and 9 and for thorough reviews containing constructive comments and valuable suggestions that helped to improve the manuscript significantly.